\documentstyle[epsfig]{aipproc}

\begin{document}
\title{Spectra of GRB 970228 from the Transient Gamma-Ray Spectrometer}

\author{D. M. Palmer$^{*\dagger}$, T.L. Cline$^{*}$, N. Gehrels$^{*}$,
K. Hurley$^{\circ}$\\
P. Kurczynski$^{*}$, N. Madden$^{\sharp}$,
R. Pehl$^{\sharp}$, R. Ramaty$^{*}$,\\\
H. Seifert$^{*\dagger}$, B.\ J.\ Teegarden$^{*}$}
\address{$^*$Goddard Space Flight Center,
$^{\dagger}$Universities Space Research Association,\\
$^{\circ}$University of California, Berkeley,
$^{\sharp}$Lawrence Berkeley Labs}

\maketitle

\def\eg    {{\it e.g.}, }
\def\gray  {\hbox{$\gamma$-ray} }
\def\grays  {\hbox{$\gamma$-rays} }
\def\etal		{{\it et al.}}
\def\ignore#1{}

\begin{abstract}
Visible afterglow counterparts have now been detected for two GRBs (970228
and 970508) but are absent, with $L_{opt}/L{\gamma}$ ratios at least two
orders of magnitude lower, for other GRBs, \eg 970828.  The causes
of this variation are unknown.  Any correspondence which could be
discovered between the \gray properties of a
GRB and its $L_{opt}/L{\gamma}$ would be useful, both in determining
the GRB mechanisms, and in allocating resources for counterpart searches
and studies.  This paper presents the  \gray spectra of GRB 970228 as
measured by the Transient Gamma-Ray Spectrometer and comments on
characteristics of this GRB compared to others that do and do not have
observable counterparts.
\end{abstract}

\section*{Introduction}

The detection of counterparts to GRBs has revolutionized the field, firmly
establishing that at least some GRBs are highly energetic objects at
cosmological distances (see numerous other papers in these
{\it Proceedings}).  As is the case with most other observable parameters of
GRBs, the luminosities of afterglow counterparts vary over a wide range
among bursts.  For example, GRB 970828 was brighter in \gray flux and
fluence than GRB 970508, yet it showed no optical counterpart, with an upper
limit more than 4 magnitudes dimmer than the peak measured for the
970508 optical transient.

By definition, all GRBs have strong \gray emission, and in most cases \gray
measurements are the only available data on which to base follow-up
observations.  With the limited resources available at short notice, it is
often necessary to select for further study only those GRBs which are most
likely to produce counterpart detections.  This is likely to cause
selection effects in the long term (if early experience suggests \eg that
only long-duration bursts have counterparts, then no counterparts to
short-duration GRBs will be sought and found).  In the short term, it is an
unfortunate reality.

GRB 970228, the first GRB to show an optical counterpart, was observed by
the Transient Gamma-Ray Spectrometer (TGRS), a 35 $\mbox{cm}^2$ germanium
spectrometer operating on the WIND spacecraft \cite{tgrs95}.  TGRS is
exposed only to the Southern ecliptic hemisphere, and so it did not see the
only other GRB which has so far yield an optical counterpart: the Northern
hemisphere GRB 970508.  TGRS spectra of other GRBs which had no comparable
optical counterpart will be the subject of a later paper.

\section*{Analysis Techniques}

\ignore{
Gamma ray spectrometers measure (with an accuracy dependent on their
detector technology) the energy deposited by each
\gray that interacts in their sensitive volume.  Some \grays pass
completely through the detector without interacting, and some \grays
deposit only part of their energy \eg by Compton scattering and then
escaping.  The solid-state germanium detector of
TGRS has excellent resolution of a few keV ({\it cf} 5-10\% for
well-designed scintillating spectrometers) and has a mean thickness of more
than 6 cm, allowing it to totally capture the energy from a large fraction
of the incident \grays (photopeak fraction).
}
\ignore{
Even with TGRS, photon spectra can be derived from the measured `counts
spectra' of deposited energy only by deconvolving with the instrumental
response function.  Since direct inversion of the instrument response tends
to be numerically unstable, a technique known as `forward convolution' is
used.  In this technique, a parameterized model is assumed for the
photon spectrum, then propagated through the detector response and compared
to the actual counts spectrum data.  The parameters of the model are
adjusted to yield the best fit of the data.  The counts spectrum is
converted to a derived photon spectrum using adjustment factors based
on the ratio of the model photon spectrum to its convolved form
\cite{tgrs95}.
}
\ignore{
Although the photon spectra obtained by forward convolution are somewhat
model-dependent, this dependency can be quantified by fitting with multiple
models and comparing the results.  Variations due to models which all fit
the data well (\eg have similar and good values of {$\chi^2$}) tend to be
small compared to the statistical errors in the data, at least for
broad-band continuum models.  The use of the models to estimate the flux of
a spectrum tends to be much less reliable, especially when the data is
available only in coarse energy bins or when the model is extrapolated
to energies where the detector loses sensitivity.  For example an
$E^{-\alpha}$ power-law spectrum has infinite flux at the high end for
$\alpha \leq 2$.
}

We use the `forward convolution' method to fit the measured,
background-subtracted `counts spectra' of deposited energy to spectral
models and to obtain photon spectra.  We use a detector response
matrix calculated by a Monte Carlo code\cite{tgrs95} (simulated with
$80^\circ$ incidence angle for this GRB).  Although this produces
model-dependent results for the photon spectra, the dependency tends to be
small for smooth continuum models over energy ranges where the instrument
has good efficiency.

\begin{figure}[t]
  \center{\vbox{\epsfig{file=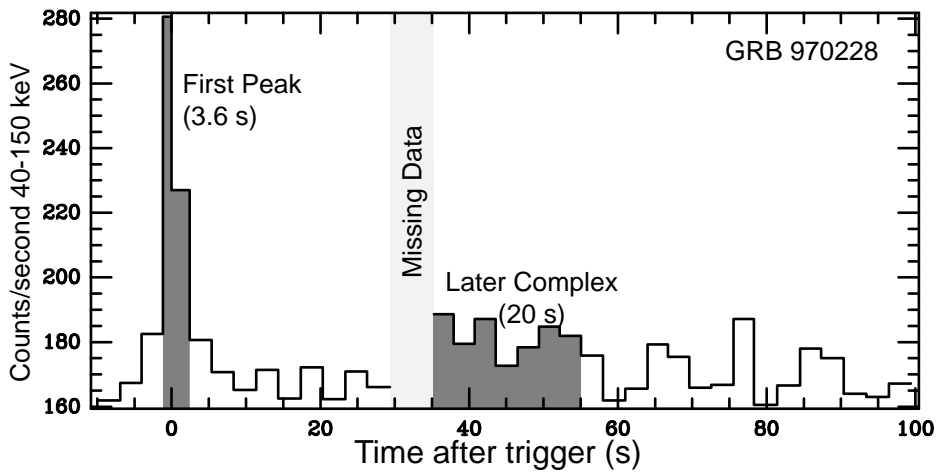}}}\par
  \caption[]{Light curve of GRB 970228 in the 40-150 keV measured-energy
range.} 
\label{figcurve}
\end{figure}

\section*{Data and Results}

Figure \ref{figcurve} shows the 40-150 keV light curve of this burst as seen
by TGRS.  The temporal resolution is set by the binning of the 0-190 keV
histogram data type used in this analysis.  Events at energies above 190
keV are recorded individually and combined with the histogram data to
produce spectra up to 8 MeV.  The typical resolution below 1 MeV is 2-4
keV.  The two intervals of this GRB which will be discussed in this paper
are marked as the `first peak' and the `later complex' on the light curve. 
The missing data before the later complex does not contain significant
flux, as verified by light curves produced using other TGRS data types and
by other instruments
\cite{hurley97}.  Observations of this burst by the instruments on the
BeppoSAX spacecraft show that the later complex consists of three softer
pulses peaking at approximately 40, 52, and 70 seconds after the trigger
time, with the 40~s pulse exceeding the amplitude of the first pulse in the
2-10 keV band
\cite{frontera97}.  The 70~s pulse is too weak in our energy band to
contribute significantly, so it is excluded from this analysis.

\begin{figure}[t]
  \centerline{\vbox{\epsfig{file=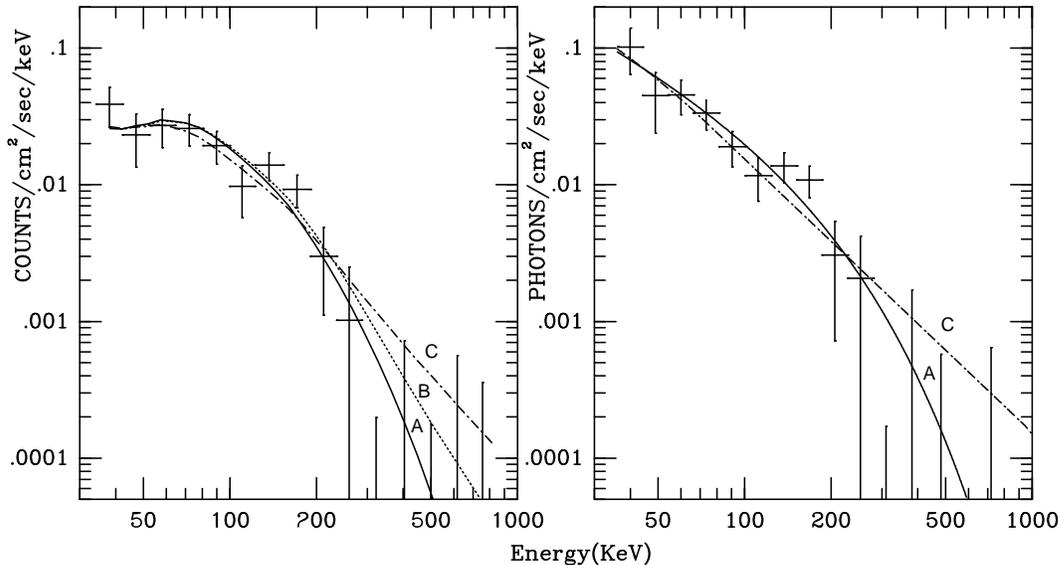, width=5.5in}}}\par
  \caption[]{Spectrum of the first peak of GRB970228.  On the left is the
TGRS counts spectrum and the best fits by 3 models.  On the right is the
derived photon spectrum.  Line
A (solid) is a good fit with an OTTB model, Line B (dotted) is a marginal
fit to a BAND model with
$\beta = -3$, Line C (dot-dashed) is an unacceptable fit to a Band model
with $\beta = -2$.  The photon spectrum is derived from the OTTB fit; a
reconstruction with the $\beta = -2$ model tends to give lower fluxes, but
the difference is small compared to the error bars.} 
\label{figpeakspect}
\end{figure}

Figure \ref{figpeakspect} shows the spectrum of 3.6~s period labeled `first
peak'.  The three model fits shown are an Optically Thin Thermal
Bremsstrahlung (OTTB) and two Band model spectra.  The Band
model\cite{bandmod} is a spectral form with a low energy $E^\alpha$ power
law, rolling over exponentially to a high energy $E^\beta$ power law.  For
the special case of $\alpha = -1$ and $\beta\rightarrow-\infty$, this form
is equivalent to the OTTB form $F(E) = E^{-1} e^{-E/T}$.  The Band model
has been quite successful in matching the spectral shapes of GRBs (within
the limits of the resolution and sensitivity of the current generation of
instruments) and has the added advantage of being a purely mathematical
form with no misleading implications about the physical processes of the
GRB.  Likewise, the use of an OTTB spectral form should not be taken to
imply that the emission process is bremsstrahlung, nor that the emitter is
optically-thin.

When the first peak spectrum is fit with a Band model with all parameters
free, $\beta$, the high-energy index, tends towards $-\infty$ and $\alpha$
takes on a value \hbox{near -1}.  This implies that an OTTB would fit this
spectrum well, and indeed it does, with parameter
$T = 119\pm^{28}_{18}$ keV giving $\chi^2 = 26.3~/~26$ d.o.f., shown as the
solid Line A in Figure \ref{figpeakspect}.

Pendleton \etal \cite{nhe} have found that pulses in GRBs can be classified
as either High-Energy or No-High-Energy  (HE or NHE) pulses, where NHE
pulses have no detectable emission above 300 keV.  A burst may be composed
purely of one type of pulse, or may be mixed.  HE pulses typically have
Band \hbox{$\beta\approx-2$}.  By constraining
$\beta$ we can set upper limits on the
strength of the high-energy emission from a spectrum.  For the first peak,
constraining the fit to $\beta\geq-3$ gives an acceptable
$\chi^2 = 28.7~/~25$ d.o.f. ($\Delta\chi^2 = +2.4$ with one additional
unconstrained parameter, shown by the dotted Line B), but $\beta\geq-2$ is
excluded with
$\chi^2 = 38.7~/~25$ d.o.f. ($\Delta\chi^2 = +12.4$, the dash-dotted Line
C).  Thus, we consider the first peak of the GRB to be an NHE pulse.

From the photon spectral models, we can determine fluences for
these GRBs.  In the 50-300 keV energy band, the fluence of this peak
during these 3.6 seconds is $1.7\pm0.1 \times 10^{-6} \mbox{ erg cm}^{-2}$,
where the error range covers all three models.  The estimated fluence for
the
\hbox{300-1000 keV} energy range is much more model dependent, with values
of $[0.24, 0.54, 1.2] \times 10^{-6} \mbox{ erg cm}^{-2}$ for the
[OTTB, $\beta=-3$, $\beta=-2$] models, respectively.  This demonstrates the
need for a good spectral model for determining the fluence of a GRB outside
a detector's sensitivity range.  (Indeed, a $\beta\geq-2$ model has infinite
energy fluence when extrapolated to high energies.)

If we make the simplifying assumption of no spectral evolution during the
analyzed 3.6 seconds of the first peak, we can pro-rate this fluence by the
light curve to find that the peak flux, over the 1.2 second first bin of
this peak, is $0.7 \times 10^{-6} \mbox{ erg cm}^{-2}\mbox{ s}^{-1}$ for the
50-300 keV band.

\begin{figure}[t]
  \center{\vbox{\epsfig{file=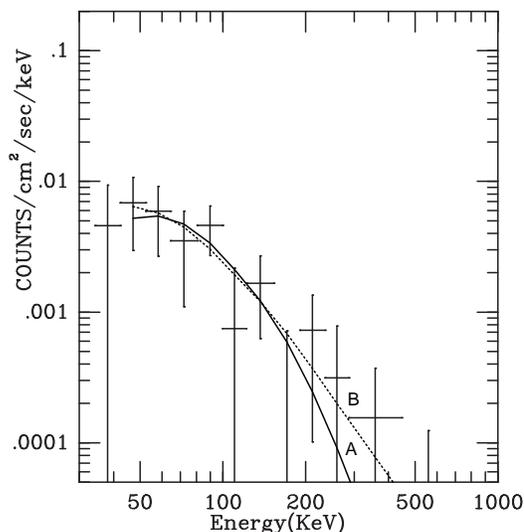, width=2.7in}}}\par
  \caption[]{Counts spectrum of the later complex of GRB970228.
The two fits are an OTTB with $T = 74$ keV (solid Line A) and a single
power law with $\alpha = -2.5$.} 
\label{figcomplexspect}
\end{figure}

Figure \ref{figcomplexspect} shows the counts spectrum of the 20 seconds we
have analyzed in the later complex.  This series of peaks has lower flux
in our energy range than the first peak, and so the measurements have more
fractional statistical error due to background counts.  An OTTB and a
single power law fit the data about equally well, with a slight advantage 
($\Delta\chi^2 = 0.8$) to the power law with $\alpha = 2.5 \pm 0.6$.  A
second power law (as in the Band model) is not required by the data at our
level of sensitivity.  The fluence for this period is
$1.4 \times10^{-6} \mbox{ erg cm}^{-2}$ for 50-300 keV, and
$[0.1, 1.0] \times 10^{-6} \mbox{ erg cm}^{-2}$ for 300-1000 with
[OTTB,  $\alpha = 2.5$] models.

\section*{Discussion}

GRB 970228, one of the two bursts known so far to have an optical
counterpart, has at least one NHE pulse.  Since NHE pulses seem to be rarer
than HE pulses (albeit with an extreme selection bias) it is natural to
wonder whether optical counterparts are produced only by NHE GRBs.  The
only other GRB with an optical counterpart, 970508, provides a
counterexample.  This GRB was seen by BATSE to be hard, with a
fluence  above 300 keV that was twice the fluence for 25-300 keV
\cite{batsecat}.

A comparison between 970228 and 970508 shows that the GRB
spectrum (or at least the existence of NHE peaks) is not immediately
successful in predicting the existence of an optical counterpart.  The
case of 970828 shows that \gray flux and fluence are not good predictors
either.  At our current state of knowledge, it seems that the most reliable
way to determine whether a GRB has an optical counterpart is simply to look.

\end{document}